# Room Temperature Electrocaloric Effect in Layered Ferroelectric CuInP$_2$S$_6$ for Solid State Refrigeration

Mengwei Si[1,4], Atanu K. Saha[1], Pai-Ying Liao[1,4], Shengjie Gao[2,4], Sabine M. Neumayer[5], Jie Jian[3], Jingkai Qin[1,4], Nina Balke[5], Haiyan Wang[3], Petro Maksymovych[5], Wenzhuo Wu[2,4], Sumeet K. Gupta[1] and Peide D. Ye[1,4,*]

[1] *School of Electrical and Computer Engineering, Purdue University, West Lafayette, Indiana 47907, United States*

[2] *School of Industrial Engineering, Purdue University, West Lafayette, Indiana 47907, United States*

[3] *School of Materials Science and Engineering, Purdue University, West Lafayette, In 47907, United States*

[4] *Birck Nanotechnology Center, Purdue University, West Lafayette, Indiana 47907, United States*

[5] *Center for Nanophase Materials Sciences, Oak Ridge National Laboratory, Bethel Valley Road, Oak Ridge, Tennessee 37831, United States*

* Address correspondence to: yep@purdue.edu (P.D.Y.)

**Abstract**

A material with reversible temperature change capability under an external electric field, known as the electrocaloric effect (ECE), has long been considered as a promising solid-state cooling solution. However, electrocaloric (EC) performance of EC materials generally is not sufficiently high for real cooling applications. As a result, exploring EC materials with high performance is of great interest and importance. Here, we report on the ECE of ferroelectric materials with van der Waals layered structure ($CuInP_2S_6$ or CIPS in this work in particular). Over 60% polarization charge change is observed within a temperature change of only 10 K at Curie temperature. Large adiabatic temperature change ($|\Delta T|$) of 3.3 K, isothermal entropy change ($|\Delta S|$) of 5.8 J $kg^{-1}$ $K^{-1}$ at $|\Delta E|$=142.0 kV $cm^{-1}$ at 315 K (above and near room temperature) are achieved, with a large EC strength ($|\Delta T|/|\Delta E|$) of 29.5 mK cm $kV^{-1}$. The ECE of CIPS is also investigated theoretically by numerical simulation and a further EC performance projection is provided.

**Keywords:** $CuInP_2S_6$, ferroelectrics, electrocaloric effect, 2D, room temperature

Electrocaloric refrigerators using electrocaloric materials are low noise, environment-friendly and can be scaled down to small dimensions, compared to the common vapor-compression refrigerators.[1–13] Electrocaloric cooling is also much easier and lower cost to realize compared to other field induced cooling techniques such as magnetocaloric and mechanocaloric cooling, because the electric field is easily to be realized and accessible. Thus, electrocaloric effect is promising for future cooling applications, especially in micro- or nano-scale such as on-chip cooling. Electrocaloric effect in ferroelectric materials is of special interest because of the large polarization change near the ferroelectric-paraelectric (FE-PE) phase transition temperature (Curie temperature, $T_C$). But it also defines the working temperature range for such ferroelectric coolers so that above but near room temperature $T_C$ is important for practical applications. Therefore, near Curie temperature as working temperature, adiabatic temperature change ($|\Delta T|$, $|\Delta T|/|\Delta E|$ as EC strength if normalized by electric field) and isothermal entropy change ($|\Delta S|$, $|\Delta S|/|\Delta E|$ if normalized by electric field) are key parameters for the performance of EC materials. One of the key challenges in realizing electrocaloric cooler is the relatively low $|\Delta T|$ and $|\Delta S|$ in current EC materials. The realization of EC cooler requires searching EC materials with high EC performance. Ferroelectric materials with van der Waals layered structure, featured with a van der Waals weak interaction between layers and being easy to form van der Waals hetero-structures, may have essential impact on the ferroelectric polarization switching and EC properties because the different out-of-plane ferroelectric polarization switching process due to the van der Waals gap. Meanwhile, the thermal transport properties of van der Waals layered materials have strongly anisotropicity in in-plane and out-of-plane directions and may provide special properties on heat absorption, heat dissipation and the design of practical ECE cooling device. Ferroelectricity in 2D materials are recently started to be studied,[14–17] but is rather rare currently because of the limited research efforts. EC materials with van der Waals heterostructures remain unexplored. Meanwhile, an insulating EC material is also required in EC refrigerator to avoid the Joule heating. CuInP$_2$S$_6$ (CIPS) has been recently explored as a 2D ferroelectric insulator with $T_C$ about 315 K and switchable polarization down to ~4 nm.[14,15,18] As a 2D ferroelectric insulator with $T_C$ above but near room temperature, CIPS can be a potential candidate for EC cooling applications.

Here, we report on the ECE on a ferroelectric insulator CIPS with van der Waals layered structure. The $T_C$ at 315 K is only slightly above human body temperature so that the material can have a broad range of practical cooling applications. Over 60% polarization change is observed with a temperature change of only 10 K. $|\Delta T|$ of 3.3 K and $|\Delta S|$ of 5.8 J kg$^{-1}$ K$^{-1}$ at $|\Delta E|=142.0$ kV cm$^{-1}$ and at 315 K are achieved, with a large EC strength ($|\Delta T|/|\Delta E|$) of 29.5 mK cm kV$^{-1}$. These representative values of CIPS suggest that ferroelectric materials with van der Waals layered structure can be competitive EC materials and is of great interest to further explore EC materials with van der Waals layered structure for potential applications in microelectronics, bio- or medical sensing, and nano-energy areas.

**Results and Discussion**

CIPS crystals were grown by solid state reaction.[18,19] Fig. 1a shows the crystal structure of CIPS from top- and side-view. It is based on a hexagonal ABC sulfur stacking, which is filled by Cu, In and P-P pairs and separated by a van der Waals gap. High-angle annular dark field STEM (HAADF-STEM) image of thin CIPS flake is shown in Fig. 1b. Distinct arrangement of atoms could be clearly identified, with the fringe space of (100) planes measured to be 0.57 nm. The corresponding selected area electron diffraction (SAED, at a 600 nm by 600 nm region) shows a set of rotational symmetry pattern with perfect hexagonal crystal structure, indicating the CIPS flake is highly single-crystallized (Fig. 1b inset). EDS analysis by SEM on CIPS flakes confirms the $CuInP_2S_6$ stoichiometry.[18] Fig. 1c illustrates the Raman spectrum of an exfoliated CIPS thin film from 4 K to 325 K. The structure of $CuInP_2S_6$ in the ferrielectric phase is in the space group Cc, point group m.[19] The primary bands in Raman study are $\nu$(P-S), $\nu$(P-P), $\nu$(S-P-S), and $\nu$(S-P-P) in the 100 to 500 cm$^{-1}$ range,[20] which shows a dramatic loss in intensity and peak broadening as the temperature is increased from 310 K to 315 K, as shown in Fig. 1d for a particular example at 373.8 cm$^{-1}$. Such measurements were repeated on multiple CIPS flakes and show similar loss in intensity and peak broadening at 315 K. Thus, temperature-dependent Raman measurements confirm a structural phase transition in CIPS at 315 K.

Temperature-dependent Piezo-response force microscopy (PFM) was studied to investigate the ferroelectricity in CIPS. Fig. 2a-c show the band excitation PFM (BE-PFM) on a fabricated CIPS capacitor using Ni as top and bottom electrodes, the area of the top electrode is 2 μm by 2 μm. Fig. 2a shows the BE-PFM images on an as-fabricated CIPS device, both up and down polarization can be observed in PFM phase image, indicating the as-fabricated CIPS device has multi-domains. DC voltage pulses of 1 s and ±6 V was then applied to the device to switch the polarization electrically. Clear phase transition can be observed after DC voltage pulses, suggesting a switchable polarization in the fabricated CIPS capacitor, as shown in the different phases (indicating different polarization directions) in Fig. 2b and 2c. The temperature-dependent ferroelectricity was further characterized by dual AC resonance tracking piezo-response force microscopy (DART-PFM). The DART-PFM phase/amplitude images and raw data of single point hysteresis loop measurements can be found in supplementary section 1. The phase and amplitude hysteresis loops were achieved by the DART-PFM at a single point in tapping mode. Fig. 2d and 2e show the phase and amplitude *versus* voltage hysteresis loop of a 0.23 μm thick CIPS flake on a Ni/SiO$_2$/Si substrate at 305 K, showing clear ferroelectric polarization switching under an external electric field (PFM phase change of ~180 degree). A 90 nm SiO$_2$ is used in between Ni and Si for better visibility of CIPS flakes. Fig. 2f shows the temperature-dependent DART-PFM phase hysteresis loop of the same CIPS flake at 300-325 K. Clear ferroelectric PFM hysteresis loops with distinct polarization switching are achieved at 300-310 K (see also Fig. S2), while no obvious phase change can be observed at and above 315 K (see also Fig. S3). The loss of ferroelectric phase transition from PFM measurement at and above 315 K, directly confirms that CIPS has a Curie temperature of ~315 K.

The electrical characteristics of CIPS devices was measured using a Ni/CIPS/Ni capacitor on top of a 90 nm $SiO_2$/Si substrate. The detailed fabrication process can be found in methods section. Polarization-voltage (P-V) measurement is used to further investigate the ferroelectric and electrocaloric properties in CIPS. The voltage-dependent P-V characteristics show a stable coercive voltage ($V_c$) *versus* sweep voltage ranges, as shown in supplementary section 2, suggesting the fabricated CIPS capacitor is highly single-crystallized, which is consistent with the TEM results in Fig. 1b. Fig. 3a shows the P-V measurement on a CIPS capacitor with CIPS thickness ($T_{FE}$) of 0.95 μm at temperature from 290 K to 330 K in 5 K step, across the Curie temperature of 315 K. A monotonic decrease of polarization *versus* temperature is observed with a peak reduction at 315 K, as shown in Fig. 3b, confirms the ferroelectric to paraelectric transition is at 315 K, which is consistent with temperature-dependent Raman spectroscopy and PFM measurements. Note that a fast over 60% polarization change is obtained within only 10 K temperature change, which may be related with the van der Waals layered structure in CIPS. The electrocaloric effect in CIPS is evaluated by indirect method.[6,21] ΔT can be calculated as $-\int_{E_1}^{E_2} \frac{1}{C\rho} T \left(\frac{\partial P}{\partial T}\right)_E dE$, where C is the heat capacity, ρ is density. ΔS can be further calculated as $\int_{E_1}^{E_2} \frac{1}{\rho} \left(\frac{\partial P}{\partial T}\right)_E dE$. As can be seen from the equations of ΔT and ΔS, the fast polarization change with respect to temperature can significantly enhance the EC strength of the EC material. The density of CIPS is 3405 kg m$^{-3}$ at 295 K.[19] The heat capacity of CIPS is 557 J K$^{-1}$ kg$^{-1}$ at 315 K.[22,23] ρ are assumed to have minor change in the temperature range of interest because the experiments were performed in a narrow temperature range between 290 K to 330 K. The temperature dependent heat capacity[23] of CIPS is used in the calculation, although only minor changes happen within the range of interest. Electrocaloric temperature change of the 0.95 μm thick CIPS is plotted in Fig. 3c. A maximum $|\Delta T|$ of 2.0 K is achieved at $|\Delta E|$ of 72.6 kV cm$^{-1}$ (maximum $|\Delta T|/|\Delta E|$ of 29.5 mK cm kV$^{-1}$ and $|\Delta T|/|\Delta V|$ of 0.31 K V$^{-1}$). The corresponding $|\Delta S|$ *versus* temperature is shown in Fig. 3d, with a maximum $|\Delta S|$ of 3.6 J kg$^{-1}$ K$^{-1}$ at $|\Delta E|$ of 72.6 kV cm$^{-1}$. It is worth to note that the absolute value of remnant polarization of CIPS thin film (~0.03-0.04 C m$^{-2}$) is rather small (about one order of magnitude smaller than common ferroelectric ceramics or ferroelectric polymers). The slope of polarization percentage change with respect to temperature is actually quite high (Supplementary section 4). If this is the intrinsic property of ferroelectric materials with van der Waals layered structure, it is possible to find a high performance 2D ferroelectric material with high remnant polarization and high EC strength. Thus, ferroelectric materials with van der Waals layered structure can be competitive EC materials and of great interest to explore.

Fig. 4 investigates the thickness dependence of ECE in CIPS. Fig. 4a show the adiabatic temperature change *versus* temperature characteristics of CIPS capacitors with CIPS thickness of 0.169 μm. A maximum $|\Delta T|$ of 3.3 K and a maximum $|\Delta S|$ of 5.8 J kg$^{-1}$ K$^{-1}$ are achieved at $|\Delta E|$ of 142.0 kV cm$^{-1}$ in the CIPS capacitor with 0.169 μm thickness. Note that the enhancement of $|\Delta T|$ and $|\Delta S|$ are because thinner CIPS film can support a higher electric field. It is also consistent with previous report that thinner CIPS has larger coercive electric field.[15] As shown in Fig. 4b, adiabatic temperature change $|\Delta T|$ *versus* electric field $|\Delta E|$ of CIPS capacitors at 315 K with different CIPS thicknesses from 0.169 μm to 1.43 μm. The slope as the actual EC strength is not improved in deep sub-μm range thin films. Fig. 4c shows the thickness dependence of EC strength in CIPS at 315 K. $|\Delta T|/|\Delta E|$ shows minor thickness dependence while the $|\Delta T|/|\Delta V|$ (Fig.

4d) is inverse proportional to the thickness as expected. A maximum $|\Delta T|/|\Delta V|$ of 1.18 K V$^{-1}$ is obtained on capacitor with 0.169 μm CIPS and at a $|\Delta E|$ of 70 kV cm$^{-1}$. Under electrical breakdown strength, ferroelectric materials with van der Waals layered structure, which have the potential to offer atomically thin films, can realize large $|\Delta T|$ change by applying very small $|\Delta V|$. The required large heat capacity can be realized by integration of a large number of such nano-coolers or nano-refrigerators. However, the leakage current in ultrathin CIPS films at elevated temperatures prevents us to reliably measure the P-V hysteresis loop and study the ECE effect. The undesired leakage current might also provide joule heating effect so that ultrathin CIPS film is not suitable for real cooling applications.

Furthermore, the EC effects in CIPS are theoretically investigated by numerical simulations. Landau-Khalatnikov equation is solved by calibrating the Landau coefficients with experimentally measured remnant polarization ($P_r$=~0.04 C m$^{-2}$) and coercive-field ($E_C$=~60 kV cm$^{-1}$) of CIPS at T=295 K. Note that the measured temperature-dependent P-V curves (Fig. 3a) show an abrupt phase transition near T=315 K. Detailed simulation methods and parameters can be found in supplementary section 3. Considering the second order phase transition with a Curie-Weiss temperature, $T_0$ (or Curie temperature, $T_C$) of 315K, the simulated temperature-dependent P-V characteristics are shown in Fig. 5a. Similarly, the simulated polarization *versus* temperature characteristics are plotted in Fig. 5b. To evaluate the EC effects, $|\Delta T|$ at different temperatures are calculated for different $|\Delta E|$ from 0 to 100 kV cm$^{-1}$ (using same method as in Fig. 3) and is shown in Fig. 5c. The simulation results (Fig. 5a-c) show good agreement with the experimental results (Fig. 3c) and suggest that in the case of second order phase transition, maximum EC temperature change occurs at 315 K. In contrast to the second order phase transition, a class of FE materials (such as BTO[24-26]) exhibit a two-step hysteresis loop change, firstly from single hysteresis loop to double hysteresis loop and then from double hysteresis loop to PE hysteresis loop. Fig. 5d shows the impact of remnant polarization on the EC performance of ferroelectric materials. It can be clearly seen that EC strength is higher with higher $P_r$. Thus, a ferroelectric material with high remnant polarization with $T_c$ above but near room temperature is preferred for high performance EC applications.

**Conclusion**

In conclusion, the electrocaloric effect on a ferroelectric material with van der Waals layered structure is investigated. Over 60% polarization change is observed with a temperature change of 10 K in CIPS. A $|\Delta T|$ of 3.3 K and $|\Delta S|$ of 5.8 J kg$^{-1}$ K$^{-1}$ at $|\Delta E|$=142.0 kV cm$^{-1}$ and at 315 K are achieved, with a large EC strength ($|\Delta T|/|\Delta E|$) of 29.5 mK cm kV$^{-1}$. The EC effect of CIPS is also investigated theoretically by numerical simulation and a further EC performance projection is provided. These results suggest the investigation of electrocaloric effects in ferroelectric materials with van der Waals layered structure is of great interest and importance for microelectronics, sensing, and nano-energy applications.

**Methods**

**CuInP$_2$S$_6$ Growth.** CIPS crystals were grown by solid state reaction. Powders of the four elements were mixed and placed in an ampoule with the same ratio of the stoichiometry (246 mg Cu, 441 mg In, 244 mg P, and 752 mg S). The ampoule was heated in vacuum at 600 °C for 2 weeks to obtain the CIPS crystals.

**Device Fabrication.** CIPS flakes were transferred to the 20 nm Ni/90 nm SiO$_2$/p+ Si substrate by scotch tape-based mechanical exfoliation. Top electrode using 20 nm nickel and gold were fabricated using electron-beam lithography, electron-beam evaporation and lift-off process.

**Material Characterization.** The HAADF-STEM were performed with FEI Talos F200x equipped with a probe corrector. This microscope was operated with an acceleration voltage of 200 kV. DART-PFM was carried out on Asylum Cypher ES and the conductive AFM tip has averaged spring constant ~5 N/m. Temperature-dependent Raman measurement was done with Montana Instruments S100-CO with a 100x 0.75NA objective integrated with a Princeton Instruments FERGIE spectrometer using a 1200 g/mm grating blazed at 550 nm. A single mode fiber coupled 532 nm laser with a <1 MHz bandwidth was used as the excitation source.

**Device Characterization.** The thickness of the CIPS device was measured using a Veeco Dimension 3100 AFM or a KLA-Tencor Alpha-Step IQ. DC electrical characterization was performed with a Keysight B1500 system. Temperature-dependent electrical data was collected with a Cascade Summit probe station.

**Acknowledgements**

This material is based upon work partly supported by the Semiconductor Research Corporation (SRC) and DARPA. J.J. and H.W. acknowledge the support from the U.S. Office of Naval Research (N00014-16-1-2465) for the TEM effort. PFM measurements were supported by the Division of Materials Science and Engineering, Basic Energy Sciences, US Department of Energy. BE-PFM experiments were conducted at the Center for Nanophase Materials Sciences, which is a DOE Office of Science User Facility. The authors gratefully acknowledge C. Wall and Montana Instruments for technical support on the temperature-dependent Raman characterization. The authors would also like to thank V. Liubachko and Y. M. Vysochanskii for identifying the heat capacity of $CuInP_2S_6$ and valuable discussions.

**Supporting Information**

Supporting Information Available: This material is available free of charge *via* the Internet at http://pubs.acs.org. Additional details for DART-PFM measurement (raw data, temperature dependence and thickness dependence) on CIPS, voltage-dependent P-V measurement, ECE simulation method, and benchmarking of EC materials are in the supporting information.

## Author contribution

P.Y.L. synthesized the $CuInP_2S_6$. M.S. did the device fabrication, electrical measurement and analysis. S.G., M.S. and W.W. measured the temperature-dependent DART-PFM. S.M.N., N.B. and P.M. performed the BE-PFM measurement. J.Q., J.J and H.W. conducted the TEM and EDS measurements. A.K.S. and S.K.G. did the numerical simulation. M.S and P.D.Y. summarized the manuscript and all authors commented on it.

## Competing financial interests statement

The authors declare no competing financial interests.

## Corresponding Author

*E-mail: yep@purdue.edu

(Uzhhorod National University, 2007).

**Figure captions**

**Figure 1.** (a) Top- and side-view of CIPS, showing an ABC sulfur stacking, filled by Cu, In and P-P pairs and separated by a van der Waals gap as a ferroelectric insulator with van der Waals layered structure. (b) HAADF-STEM image of thin CIPS flake viewed along [001] axis and the corresponding SAED pattern. (c) Temperature-dependent Raman spectroscopy on CIPS thin film. (d) Temperature-dependent Raman peak intensity at 373.8 $cm^{-1}$, showing a large decrease in Raman intensity at 315 K.

**Figure 2.** (a) BE-PFM amplitude and phase of as-fabricated CIPS capacitor with 20 nm Ni as top electrode on a $Ni/SiO_2/Si$ substrate, measured at room temperature. The area of the top electrode is 2 μm by 2 μm. (b) BE-PFM amplitude and phase of the same device after 6 V voltage pulse for 1 s. (c) BE-PFM amplitude and phase of the same device after -6 V voltage pulse for 1 s. Stable polarization switching is achieved upon application of DC voltage pulses of 1 s duration and ±6 V amplitude. (d) Phase and (e) amplitude *versus* voltage by DART-PFM hysteresis loop of a 0.23 μm thick CIPS flake without 20 nm Ni top electrode on a $Ni/SiO_2/Si$ substrate at 305 K, showing clear ferroelectric polarization switching under external electric field. (f) Temperature-dependent DART-PFM phase hysteresis loop of the same CIPS flake at 300-325 K. The loss of ferroelectric phase transition since 315 K suggesting a ferroelectric Curie temperature of 315 K.

**Figure 3.** (a) Polarization *versus* voltage characteristics measured at different temperatures of a CIPS capacitor with CIPS thickness of 0.95 μm. (b) Polarization *versus* temperature at different voltage biases, extracted from (a). (c) Adiabatic temperature change |ΔT| *versus* temperature at different |ΔE|. A maximum |ΔT| of 2.0 K is achieved at |ΔE| of 72.6 kV $cm^{-1}$ and |ΔV| of 6.9 V (Maximum |ΔT|/|ΔE| of 29.5 mK cm $kV^{-1}$ and |ΔT|/|ΔV| of 0.31 K $V^{-1}$). (d) Isothermal entropy change |ΔS| *versus* temperature at different |ΔE|, with a maximum |ΔS| of 3.6 J $kg^{-1}$ $K^{-1}$ at |ΔE| of 72.6 kV $cm^{-1}$.

**Figure 4.** (a) Adiabatic temperature change $|\Delta T|$ *versus* temperature at different $|\Delta E|$ of a CIPS capacitor with CIPS thickness of 0.169 μm. A maximum $|\Delta T|$ of 3.3 K and A maximum $|\Delta S|$ of 5.8 J kg$^{-1}$ K$^{-1}$ are achieved at $|\Delta E|$ of 142.0 kV cm$^{-1}$. (b) Adiabatic temperature change $|\Delta T|$ *versus* electric field $|\Delta E|$ of CIPS capacitors with different CIPS thicknesses. (c) Electrocaloric strength $|\Delta T|/|\Delta E|$ of CIPS capacitors with different CIPS thicknesses. (d) Normalized adiabatic temperature change with respect to voltage of CIPS capacitors with different CIPS thicknesses.

**Figure 5.** (a) Polarization *versus* voltage characteristics considering second order phase transition at temperature from 290 K to 335 K in 5 K step. (b) Polarization *versus* temperature at different voltage biases, extracted from (a). (c) EC temperature change $|\Delta T|$ *versus* temperature at different $|\Delta E|$. (d) The impact of remnant polarization on EC strength.

**Figure 1.**

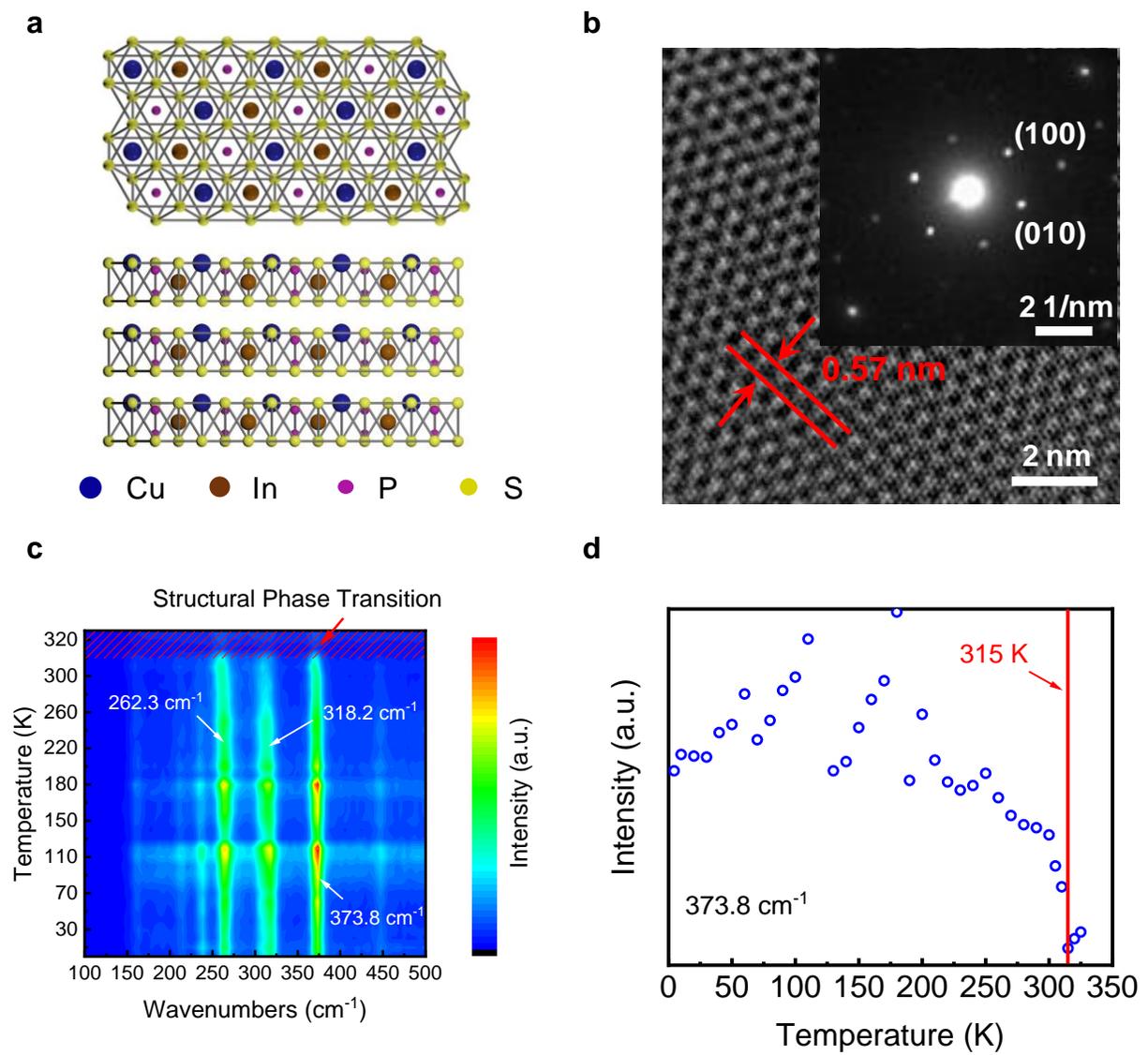

**Figure 2.**

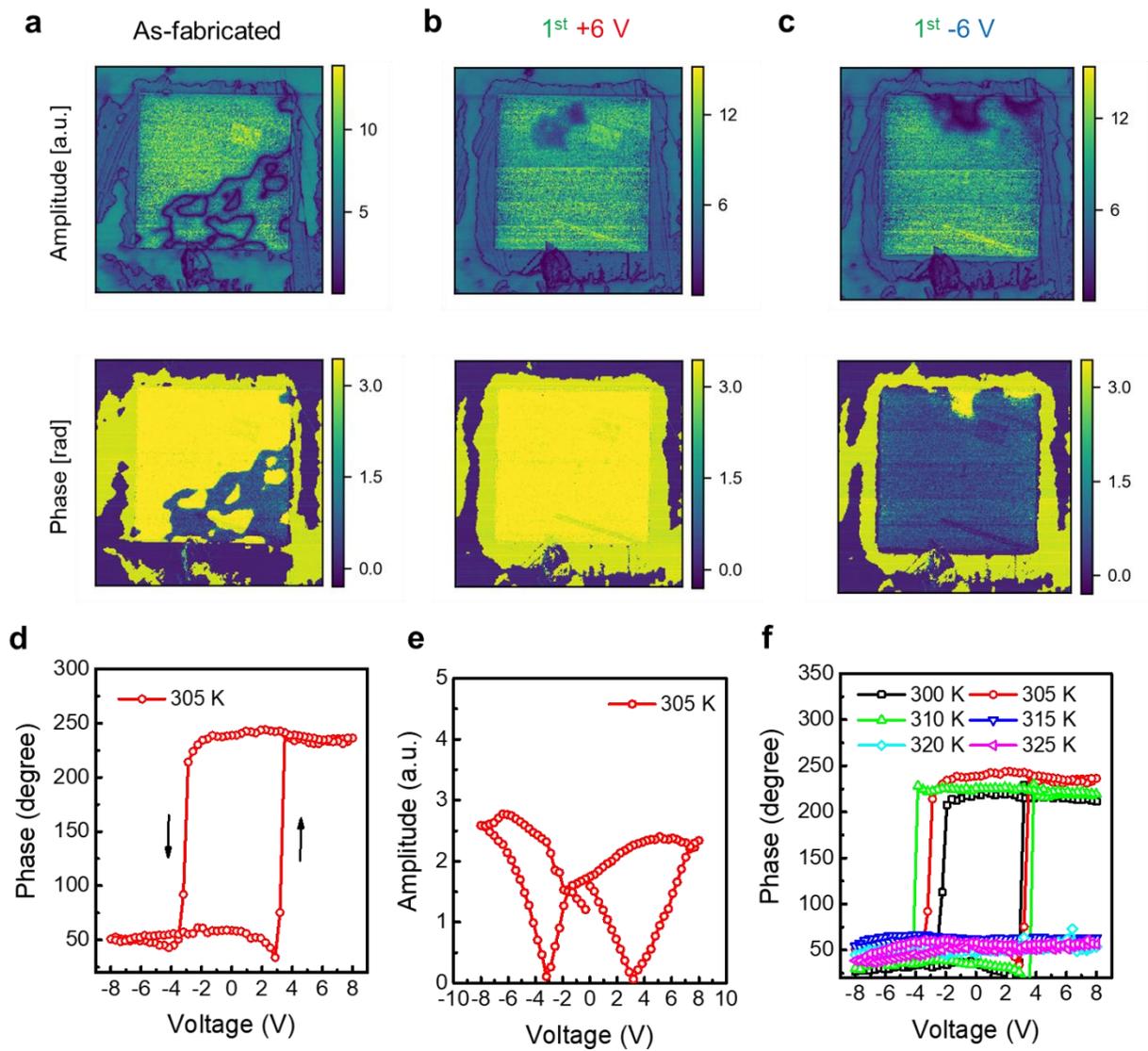

**Figure 3.**

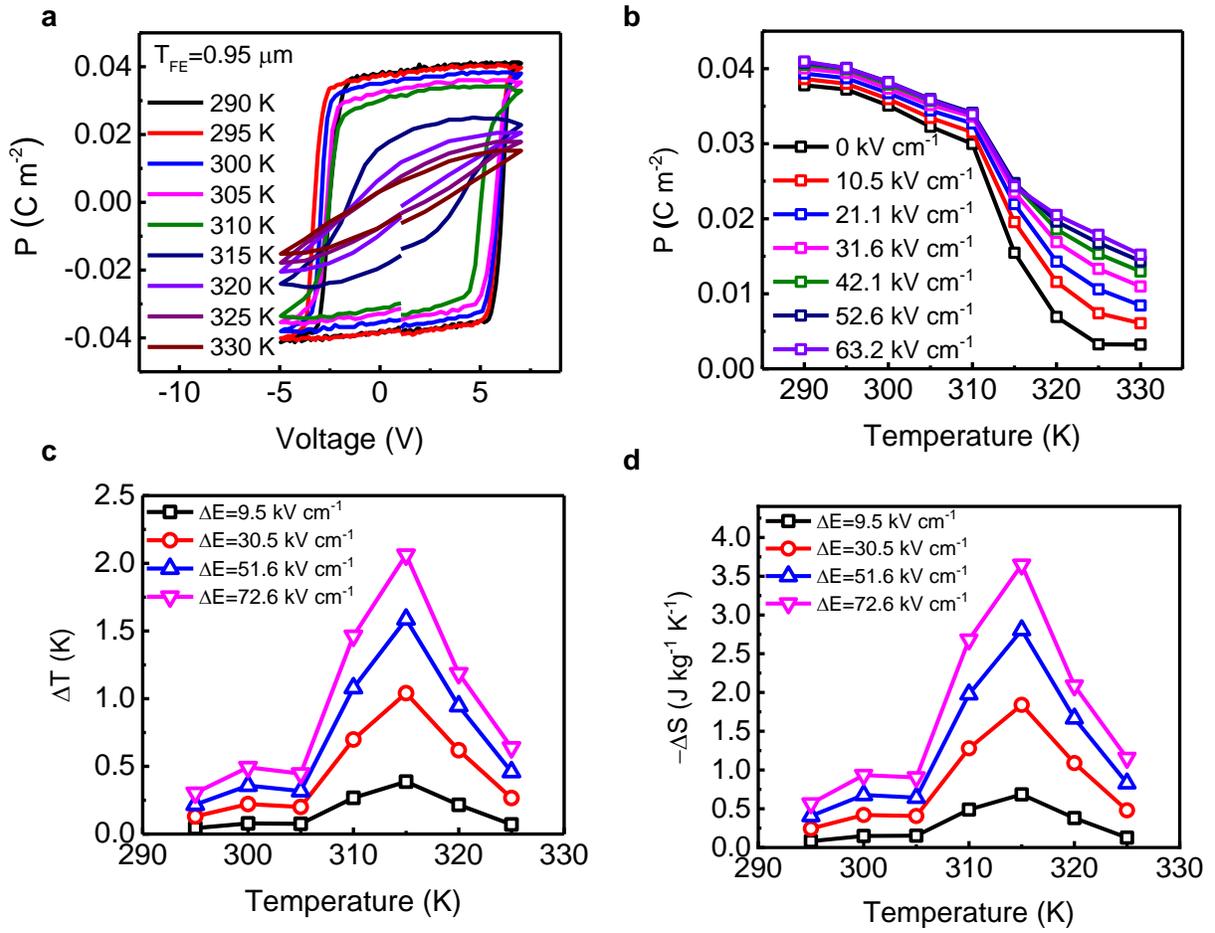

**Figure 4.**

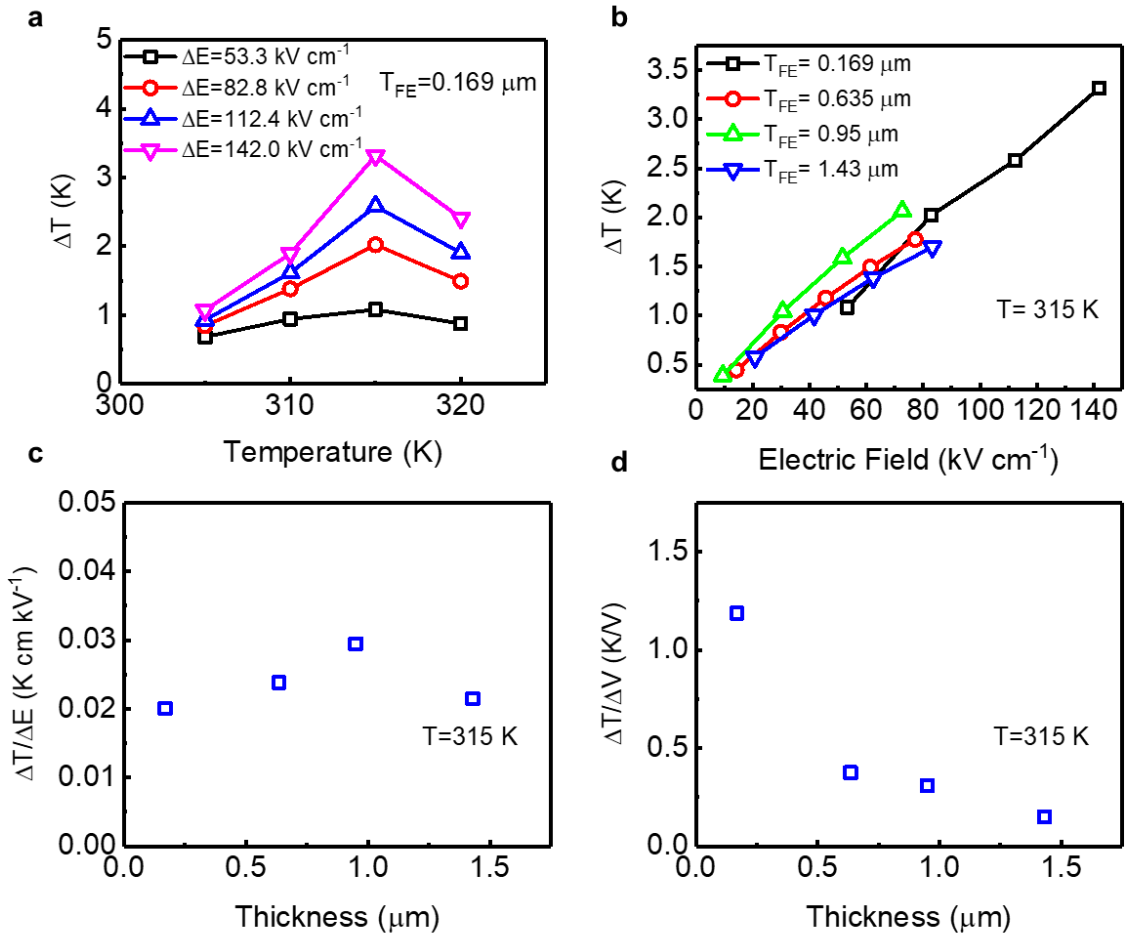

**Figure 5.**

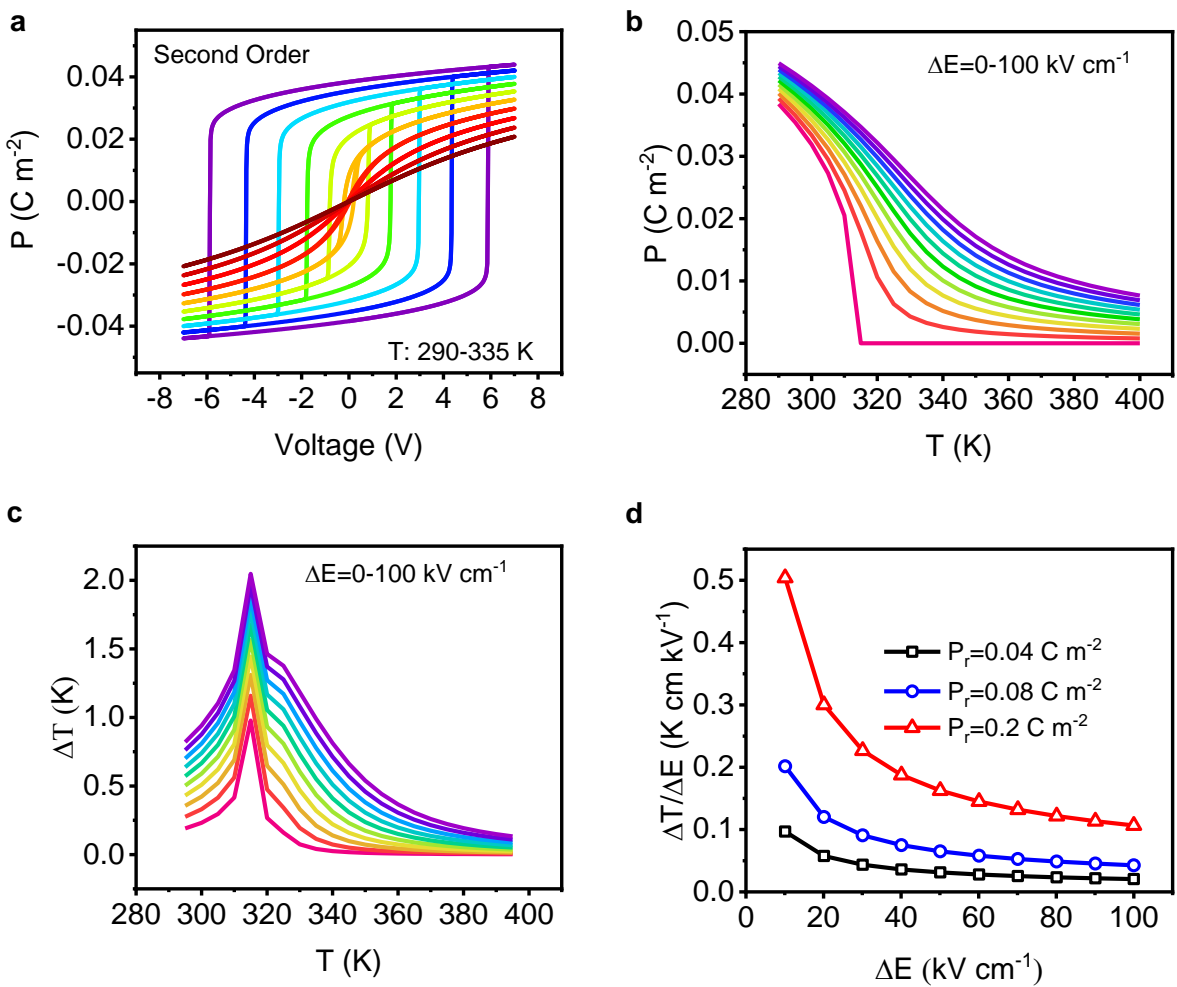

**TOC.**

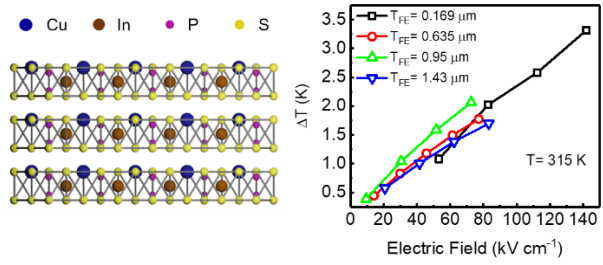

# Room Temperature Electrocaloric Effect in Layered Ferroelectric CuInP$_2$S$_6$ for Solid State Refrigeration

Mengwei Si[1,4], Atanu K. Saha[1], Pai-Ying Liao[1,4], Shengjie Gao[2,4], Sabine M. Neumayer[5], Jie Jian[3], Jingkai Qin[1,4], Nina Balke[5], Haiyan Wang[3], Petro Maksymovych[5], Wenzhuo Wu[2,4], Sumeet K. Gupta[1] and Peide D. Ye[1,4,*]

[1] *School of Electrical and Computer Engineering, Purdue University, West Lafayette, Indiana 47907, United States*

[2] *School of Industrial Engineering, Purdue University, West Lafayette, Indiana 47907, United States*

[3] *School of Materials Science and Engineering, Purdue University, West Lafayette, In 47907, United States*

[4] *Birck Nanotechnology Center, Purdue University, West Lafayette, Indiana 47907, United States*

[5] *Center for Nanophase Materials Sciences, Oak Ridge National Laboratory, Bethel Valley Road, Oak Ridge, Tennessee 37831, United States*

\* Address correspondence to: yep@purdue.edu (P.D.Y.)

## 1. DART-PFM measurement on CIPS thin film

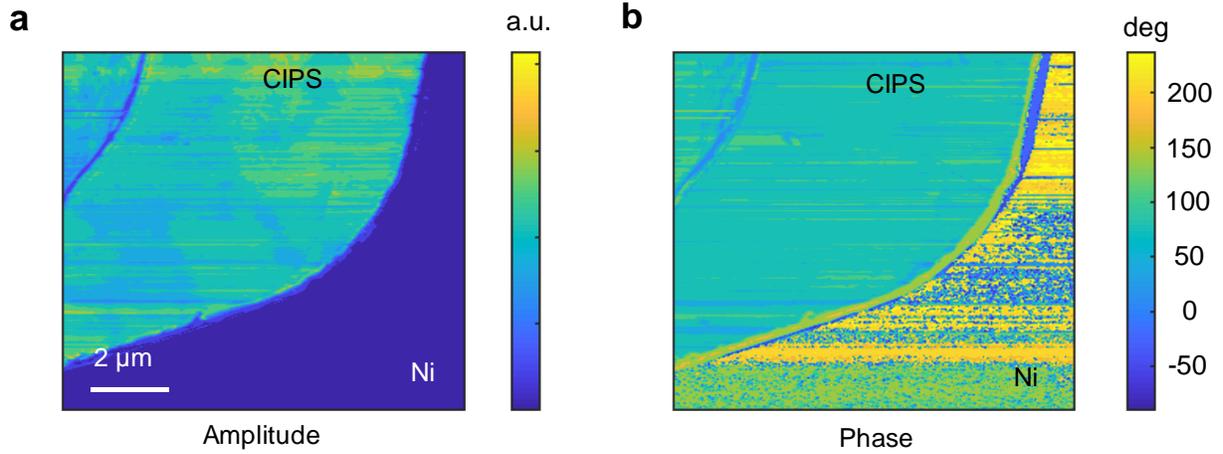

**Figure S1.** (a) PFM amplitude and (b) PFM phase images of CIPS flake on a Ni/SiO$_2$/Si substrate by DART-PFM, showing clear piezoelectric response.

Fig. S1(a) and S1(b) show the phase and amplitude by DART-PFM, suggesting a clear piezoelectric response in CIPS thin film. Fig. S2 shows the raw data of single point DART-PFM hysteresis loop measurement at 310 K. Fig. S2(a) shows the applied voltage biases versus time. Three cycles of triangular voltage waves are applied with both on field and off field PFM measurements, as shown in Fig. S2(b). Fig. S2(c)-(d) show the PFM amplitude, phase1 and phase2 signals. A clear ferroelectric polarization switching can be seen on both phase1 and phase2 signals, at both on field and off field. The raw data itself confirms the ferroelectricity of CIPS at 310 K. Fig. S3(a)-(c) show the PFM amplitude, phase1 and phase2 signals of CIPS measured at 325 K. The biases and time sequences are same as in Fig. S2(a). No ferroelectric phase switching can be seen on both phase1 and phase2 signals at off field, suggesting the loss of ferroelectricity of CIPS at 325 K. Further thickness-dependent PFM measurements are performed at room temperature to explore the scaling property of ferroelectric CIPS. As shown in the thickness-dependent PFM measurements on CIPS in Fig. S4 from 140 nm down to 15 nm, clear ferroelectric hysteresis loop can be achieved down to 40 nm, suggesting the stable ferroelectricity in CIPS down to tens of nm at room temperature.

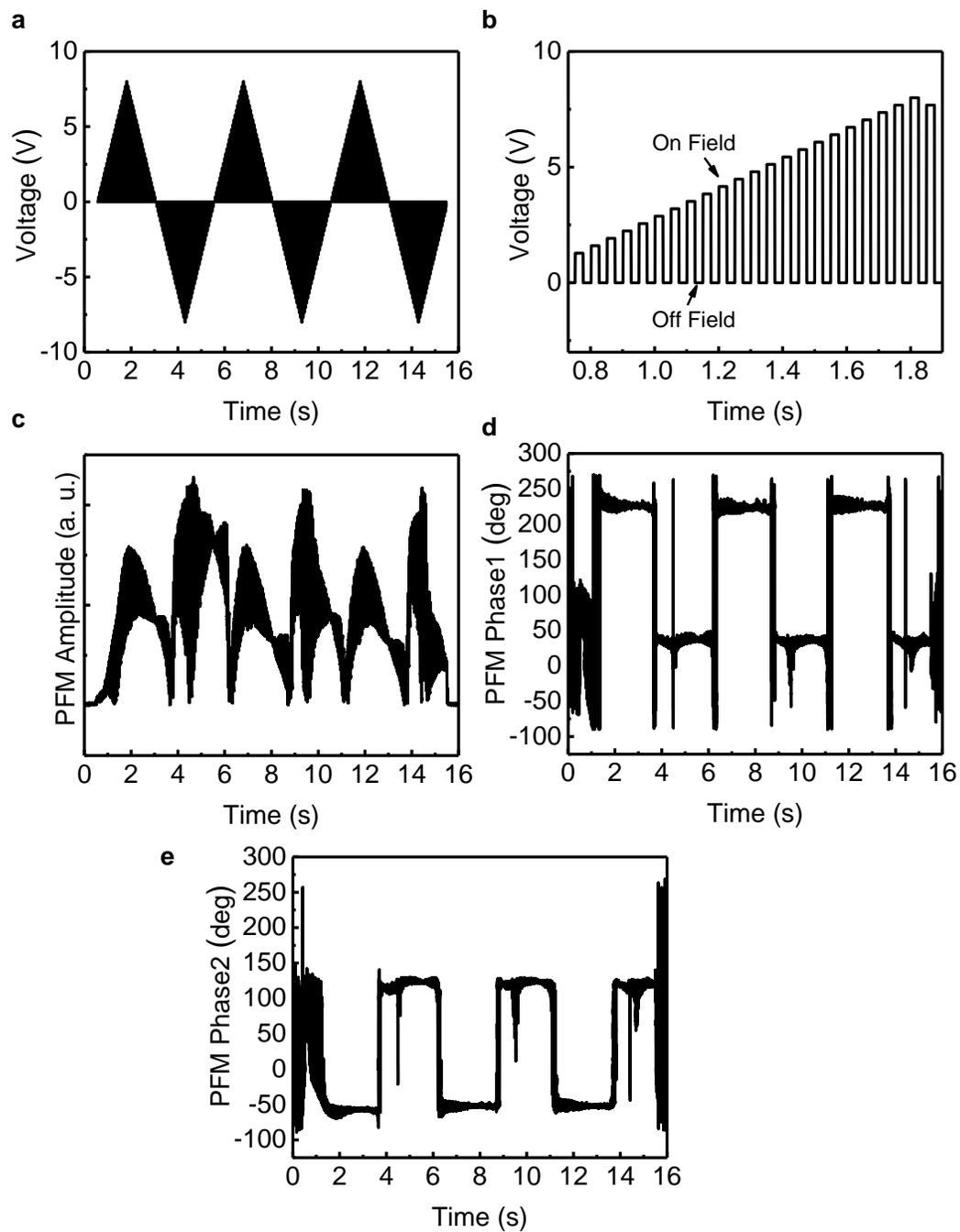

**Figure S2.** Raw data of single point DART-PFM hysteresis loop measurements at 310 K. (a) Bias, (b) zoom-in plot of (a), showing on field and off field measurements, (c) amplitude, (d) phase1, and (e) phase2.

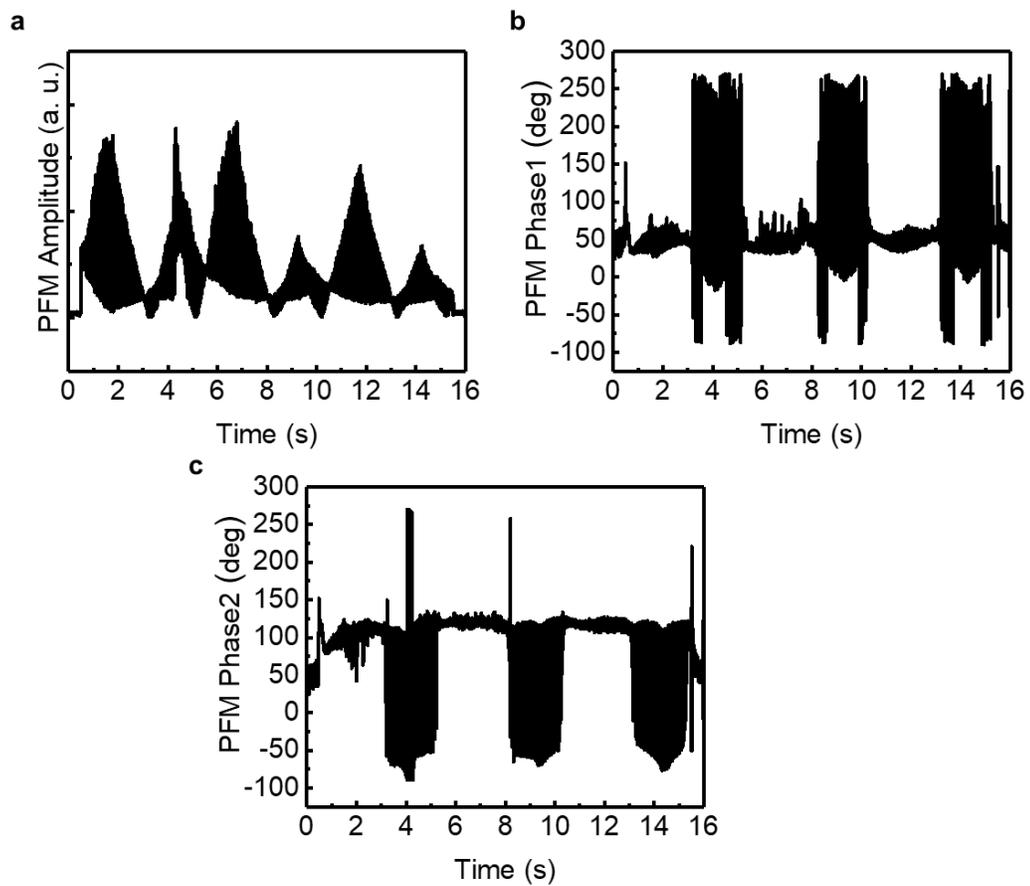

**Figure S3.** Raw data of single point DART-PFM hysteresis loop measurements at 325 K. (a) Amplitude, (b) phase1, and (c) phase2.

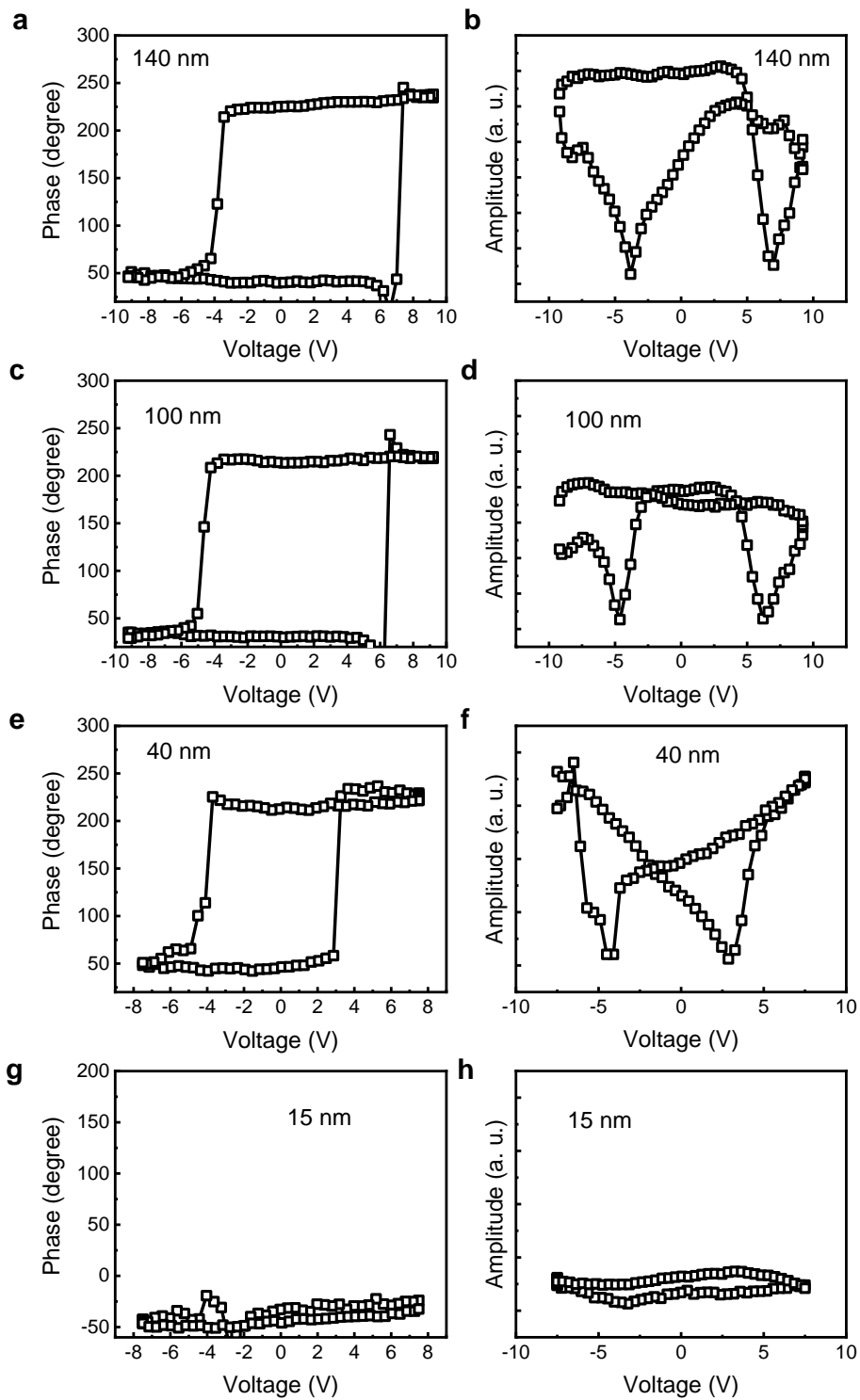

**Figure S4.** Thickness dependent PFM measurements on CIPS at room temperature. Phase versus voltage hysteresis loop at (a) 140 nm, (c) 100 nm, (e) 40 nm and (g) 15 nm. Amplitude versus voltage hysteresis loop at (b) 140 nm, (d) 100 nm, (f) 40 nm and (h) 15 nm.

## 2. Voltage-dependent ferroelectricity and ferroelectric retention

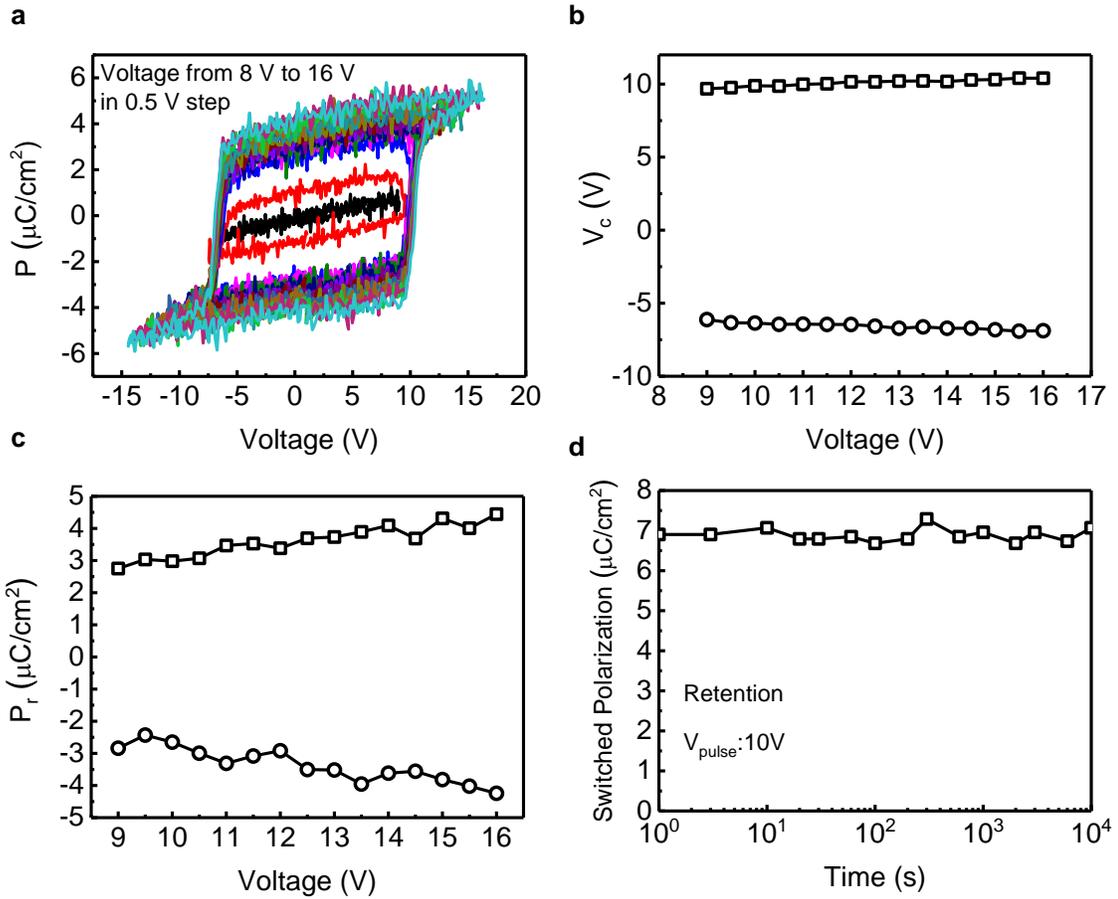

**Figure S5.** (a) P-V hysteresis loop of CIPS capacitor at different voltage sweep ranges. (b) Coercive voltage and (c) remnant polarization versus sweep voltages. (d) Switched polarization versus time in ferroelectric retention measurement, showing retention-free up to $10^4$ s.

Fig. S5(a) shows voltage-dependent P-V hysteresis loop of CIPS capacitor. Fig. S5(b) shows the coercive voltage ($V_c$) versus maximum applied voltages. Fig. S5(c) shows the remnant polarization ($P_r$) versus maximum applied voltages. Fig. S5(d) shows the polarization retention measurement of CIPS capacitor. No polarization retention is observed up to $10^4$ s.

### 3. Modelling and Calibration:

To simulate the electrocaloric effect in CIPS, we solve the temperature ($T$) dependent Landau-Khalatnikov equation (eqn. (1)) that captures the relation among polarization, temperature and applied electric field.

$$E - \rho_{FE}\frac{dP}{dt} = \alpha_0(T - T_0)P + \beta P^3 + \gamma P^5 \qquad (1)$$

Here, $P$ is the polarization, $E$ is the applied electric field, $\rho_{FE}$ is the viscosity coefficient and $t$ is time. $\alpha_0$, $\beta$ and $\gamma$ are Landau coefficients and $T_0$ is the Curie-Weiss temperature. According to eqn. (1), the temperature driven phase transition depends on the sign of $\beta$. For $\beta>0$, eqn. (1) can capture the second order phase transition and if $\beta<0$, then the equation corresponds to the first order phase transition. In case of second order phase transition, the Curie temperature ($T_C$) is the same as $T_0$. At $T=T_C=T_0$, the material changes its phase from FE to PE.

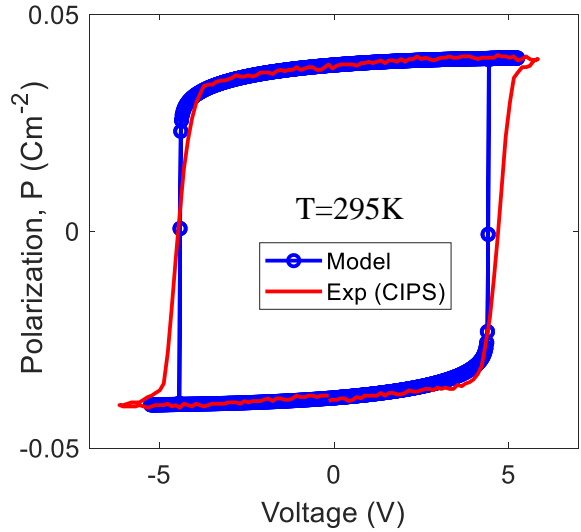

**Figure S6.** Experimentally measured and simulated P-V hysteresis loop of CIPS capacitor at T=295 K.

We calibrate the Landau coefficients by fitting the simulated polarization versus voltage (P-V) characteristics with the measured P-V characteristics for T=295 K (Fig. S6). As the experimentally measured temperature dependent P-V characteristics suggest second order phase transition in CIPS, therefore, we assume $\beta>0$, in our calibration. The corresponding Landau coefficients we get from the calibration are presented in Table SI. Then, the electrocaloric temperature change can be calculated using eqn. (2) based on the simulated temperature dependent P-V characteristics, where C is the heat capacity, $\rho$ is the mass density.

$$\Delta T = - \int_{E_1}^{E_2} \frac{1}{C\rho} T \left(\frac{\partial P}{\partial T}\right)_E dE \tag{2}$$

Table - SI: Landau coefficients considering the second order phase transition (CIPS, T=295 K, $T_0$=315 K)

| $\alpha_0(T-T_0)$ | $-3.52\times10^8$ m/F |
|---|---|
| $\beta$ | $1.38\times10^{11}$ m$^5$/F/C$^2$ |
| $\gamma$ | $6.81\times10^{13}$ m$^9$/F/C$^4$ |

## 4. Benchmarking of EC Materials:

Table SII shows the performance of selected EC materials according to Ref. 1. CIPS exhibits comparable EC performance with other 3D EC materials. Note that comparing with most thin film ferroelectric materials, such as thin film PZT[4], PMN-PT[6], SBT[7], P(VDF-TrFE)[8], the EC strength ($|\Delta T|/|\Delta E|$) of CIPS is quite high.

Note that $\Delta T$ is proportional to $\partial P/\partial T$ according to eqn. (2). $\partial P/\partial T$ can be further estimated as $P_r \partial P_\% /\partial T$, where $P_r$ is the remnant polarization and $\partial P_\% /\partial T$ is the slope of polarization percentage change. This means if CIPS has a remnant polarization of ~30-40 $\mu C/cm^2$ (similar to conventional ferroelectric materials like PZT) while maintaining the same slope of polarization percentage change, a $|\Delta T|$ of over 30 K can be achieved. This is a very high number comparing with other EC materials.

Table - SII: Performance of Selected EC Materials

| EC Material | T (K) | $|\Delta T|$ (K) | $|\Delta E|$ (kV cm$^{-1}$) | $|\Delta T|/|\Delta E|$ (mK cm kV$^{-1}$) | Ref. |
|---|---|---|---|---|---|
| $KH_2PO_4$ | 123 | 1.0 | 10 | 100 | 2 |
| $BaTiO_3$ | 397 | 0.9 | 4 | 225 | 3 |
| $PbZr_{0.95}Ti_{0.05}O_3$ | 499 | 12 | 480 | 25 | 4 |
| $Pb_{0.8}Ba_{0.2}ZrO_3$ | 290 | 45 | 598 | 75.3 | 5 |
| 0.9PMN-0.1PT | 348 | 5.0 | 895 | 5.6 | 6 |
| $SrBi_2Ta_2O_9$ | 565 | 4.9 | 600 | 8.2 | 7 |
| P(VDF-TrFE) | 323 | 28 | 1800 | 15.6 | 8 |
| CIPS (This work) | 315 | 3.3 | 142 | 29.5 | |